\documentclass[12pt, draftclsnofoot, onecolumn]{IEEEtran}
\IEEEoverridecommandlockouts

\usepackage{ifpdf}

  \usepackage[dvips]{graphicx}
  \graphicspath{{../eps/}}
  \DeclareGraphicsExtensions{.eps}

\usepackage{cite}
\usepackage{cancel}
\usepackage[normalem]{ulem}
\usepackage{algorithm}
\usepackage{algorithmic}
\usepackage[cmex10]{amsmath}
\usepackage{mathtools}
\usepackage{amsthm}
\usepackage{amssymb}
\usepackage{balance}
\usepackage{eqparbox}
\usepackage{bbm, flushend}
\usepackage{mathtools}
\usepackage[colorinlistoftodos]{todonotes}
\usepackage{color}
\newcommand{\bseq}{\begin{subequations}}
\newcommand{\eseq}{\end{subequations}}
\newcommand{\baln}{\begin{align}}
\newcommand{\ealn}{\end{align}}
\newcommand{\balnd}{\begin{aligned}}
\newcommand{\ealnd}{\end{aligned}}
\newcommand{\beq}{\begin{equation}}
\newcommand{\eeq}{\end{equation}}
\newcommand{\beqn}{\begin{eqnarray}}
\newcommand{\eeqn}{\end{eqnarray}}
\newcommand{\beqno}{\begin{eqnarray*}}
\newcommand{\eeqno}{\end{eqnarray*}}
\newcommand{\bma}{\begin{displaymath}}
\newcommand{\ema}{\end{displaymath}}
\newcommand{\bnu}{\begin{enumerate}}
\newcommand{\enu}{\end{enumerate}}
\newcommand{\bce}{\begin{center}}
\newcommand{\ece}{\end{center}}
\newcommand{\btb}{\begin{tabular}}
\newcommand{\etb}{\end{tabular}}
\newcommand{\ba}{\begin{array}}
\newcommand{\ea}{\end{array}}
\newcommand{\sdfrac}[2]{\mbox{\small$\displaystyle\frac{#1}{#2}$}}

\begin{document}

\title{{Joint Computation Offloading, SFC Placement, and Resource Allocation for Multi-Site MECs}}

\author{Phuong-Duy Nguyen and Long Bao Le \\
}

\maketitle

\vspace{-1.0cm}

\begin{abstract}
Network function Virtualization (NFV) and Mobile Edge Computing (MEC) are promising 5G technologies 
to support resource-demanding mobile applications. In NFV, one must process the 
service function chain (SFC) in which a set of network functions 
must be executed in a specific order. Moreover, the MEC technology 
enables computation offloading of service requests from mobile users to remote servers 
to potentially reduce energy consumption and processing delay for the mobile application. 
This paper considers the optimization of the
computation offloading, resource allocation, and SFC placement in the multi-site MEC system. Our
 design objective is to minimize the weighted normalized energy consumption and computing cost
subject to the maximum tolerable delay constraint. To solve the underlying mixed integer and non-linear  
optimization problem, we employ the decomposition approach where we iteratively optimize the computation offloading, 
SFC placement and computing resource allocation to obtain an efficient solution. Numerical results show 
the impacts of different parameters on the system performance and the superior performance of the proposed algorithm
compared to benchmarking algorithms. 
\end{abstract}


\IEEEpeerreviewmaketitle

\section{Introduction}

The proliferation of smartphones over the last decade has stimulated the emergence of many resource-demanding mobile applications
such as video gaming, virtual/augmented reality. The limited computation and battery capacity of mobile devices 
have become bottleneck for the deployments of many emerging mobile applications.
The MEC has been considered a potential solution to these problems where heavy computation
and processing tasks can be offloaded from mobile users to a MEC server for execution \cite{ETSI_2014}.

MEC servers can be deployed at radio base stations (BSs), which allows to process large computation tasks at the network edge.
The MEC technology, therefore, helps reduce application latency and energy consumption which 
improves the users' quality of experience \cite{mao2017survey}. Moreover, employment of NFV in the software defined networking (SDN)
based 5G wireless networks allows mobile application functions to run virtual machines or containers \cite{mehraghdam2014specifying}
where VNFs associated with a particular application can be represented by an execution graph through a process called SFC.
The dynamic deployment of VNFs from many mobile applications requires to address several challenging problems: (i) VNFs' \textit{placement}
to determine a physical host running each VNF, and (ii)  computing resource \textit{allocation} to execute the VNFs at
the assigned hosts. Joint design of SFC placement and resource allocation across multiple clouds is an important research
 problem \cite{bhamare2017optimal}.
In the 5G networks, the NVFs are originated from an application of a mobile user; therefore, one must decide whether to 
execute these NVFs on the mobile device or offloaded to remote servers for execution.
This offloading incurs communication delay and energy consumption, which must be taken into
account in the offloading decision. 

Several design aspects of MEC have been studied in the literature. Joint optimization of offloading decision and resource allocations for delay-sensitive tasks is addressed in \cite{zhang2018joint}. The dynamic voltage frequency scaling (DVFS) technique employed for 
energy saving of mobile devices is explored in \cite{dinh2017offloading}. Different approaches have been taken to
address the computation offloading design including heuristic mechanisms \cite{lyu2016multiuser}, dynamic programming 
 \cite{lyuwork}, and distributed computation replication \cite{guo2018energy}. However, joint design of native application chaining 
structure \cite{agarwal2018joint}, computation offloading, and resource allocation leveraging the collaboration among servers in the multi-site MEC system \cite{nguyen2019computation}. 
Our current paper fills this gap in the existing literature.

\begin{figure}
\centering
\includegraphics[width=0.7\columnwidth]{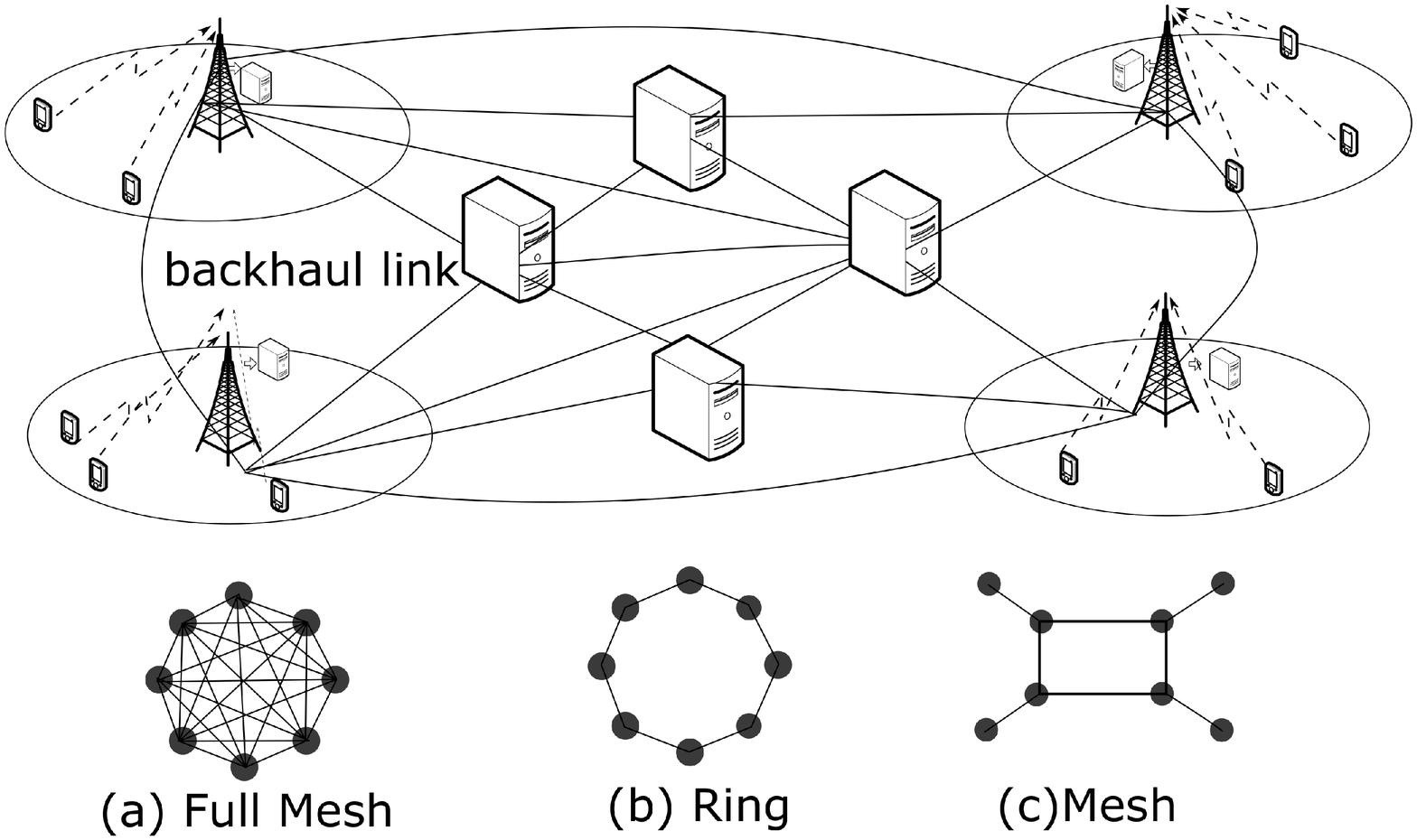}
\caption{Multi-server MEC system with different backhaul topologies}
\label{fig-MultiFog}
\vspace{-5mm}
\end{figure}

In the 5G wireless system, the edge servers deployed at individual BSs may have limited resources or lack
certain service libraries to execute underlying applications. Collaborations among edge/cloud servers, as illustrated in Fig.~\ref{fig-MultiFog},
by offloading computing load of different VNFs in the SFC using backhaul links allow efficient
execution of the underlying applications. In this paper, we consider such a multi-server MEC system
and our design jointly optimizes the offloading, placement of VNFs and computing resource allocation
to minimize the weighted sum of normalized mobile energy consumption and 
computation cost considering constraints on the maximum execution latency and maximum computing resources at the 
servers. We propose an efficient algorithm to solve this challenging problem by using the decomposition
approach and show the efficacy of this design via extensive numerical studies. 


The rest of this paper is organized as follows. Section~II presents the system model. Section III describes
 the proposed design and algorithms. Section IV evaluates the performance of our design
followed by the conclusion in Section V.

\section{System Model} \label{sec:SM}

\subsection{MEC System and Backhaul Network Models}

Consider a multi-server MEC system with several computation servers (CoSs) denoted by
the server set $\mathcal{V}$. Moreover, we assume that one edge server is deployed
at each multi-antenna base station (BS) and the set of these edge servers $\mathcal{S}$ is a subset of $\mathcal{V}$
(i.e., $\mathcal{S} \subseteq \mathcal{V}$). 
Further, each combined BS~$s$ and its co-located CoS provide both wireless communication and computing services to
mobile users (MUs) inside its coverage. For convenience, we use $\mathcal{S}$ to refer to the set of BSs and the set of associated
CoSs. Let $\mathcal{K}_s$ denote the set of MUs served by BS $s \in \mathcal{S}$.
We assume that each CoS $v$ has a limited computing capacity represented by the maximum clock speed $\bar{F}_v$ (CPU cycles per second).
For brevity, we refer to MU $k$ associated with BS $s$ as MU $(k, s)$ in the following.

We further assume that the CoSs are inter-connected by the backhaul network
where the computation load from one server can be offloaded to other one-hop-away servers
for execution. We model this backhaul network as a directed graph where the set of
CoSs $\mathcal{V}$ correspond to the nodes  and the set of backhaul
links corresponds to the set of (directed) edges in the graph. 
With this graph model, each vertex/CoS $v$ is connected with and can receive data from
 a set of the vertices adjacent to it, called in-neighbor vertex
set $\mathcal{L}^{{I}}_v$.
Therefore, any CoSs in $\mathcal{L}^{{I}}_v$ can offload their computation load to CoS $v$.
For any two connected CoSs $x$ and CoS $y$, we assume the data transmission delay
over the corresponding backhaul link is approximately equal to its connection setup time $\delta^{cm}_{xy}>0$
(i.e., backhaul transmission rate is very high).
To capture the connectivity of the CoSs, we introduce the binary parameters $\mathbbm{e}_{xy} = \mathbbm{1}_{\{\delta^{cm}_{xy}>0\}} $, where $\mathbbm{1}_{(x)}$ is the indicator function, equal 1 if there exists a connection between CoSs $x$ and $y$ and 0, otherwise.



\subsection{Service Function Chain (SFC) Model}


Let $\mathfrak{F}$ denote the set of all possible network functions. 
We assume that each CoS $v$ provides services to execute the subset of functions 
$\mathfrak{F}_v{\subseteq}\mathfrak{F}$. 
Each MU $k$ at BS $s$ is assumed to run an application whose computation load
can be decomposed into the set of service requests and their corresponding network functions.
Moreover, each request can be executed locally and/or at the remote CoSs (via offloading).
Specifically, the network functions of each request can be represented
by an ordered function set, called service function chain (SFC), where the order
of this set represents the execution order of the corresponding functions. 

\begin{figure}
\centering
\includegraphics[width=0.85\columnwidth]{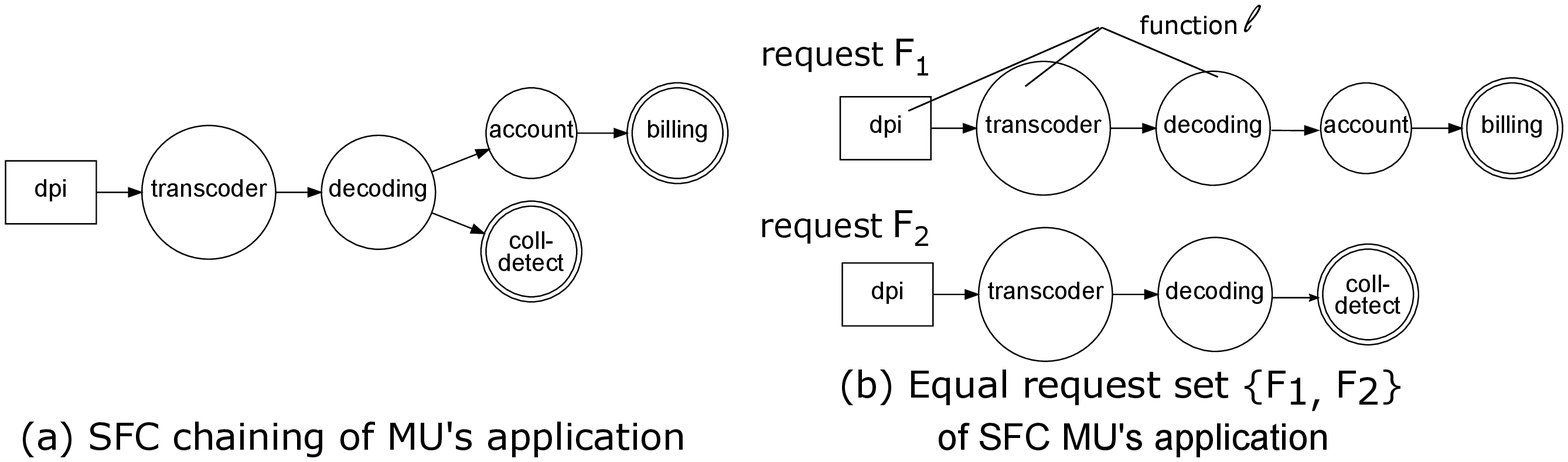}
\caption{Service function chains of mobile application}
\label{fig-SFC}
\vspace{-5mm}
\end{figure}

These request/function models are illustrated in Figure \ref{fig-SFC}.
In particular, each MU $k$ in BS $s$ has a set of service requests in the set $R_{k,s}$. Each 
request $r{\in}R_{k,s}$ corresponds to an ordered set of network functions $F_r \subseteq F$,
which must be placed and executed at the MU or some CoSs. Moreover, each function has a 
particular amount of input data (e.g., a video file) to be processed and the execution of
the function produces an amount of output data. Let $\xi_{k,s}^{r,l}$ represent the ratio
between the amount of output data after executing function $l$ 
and the original input data of request $r$ associated with MU $k$ of BS $s$. 
The parameter $\xi_{k,s}^{r,l}$ will depend on the data output/input ratios of all functions
executed before function $l$ in the SFC. 



\subsection{Computation Offloading Models}


We assume that MU $k$ at BS $s$ needs to run an application with the request set $R_{k,s}$ 
and the total amount of input data is  $\bar{u}_{k,s}$~(bits).
Moreover, each request $r$ from this application has to process
a fraction $\zeta_r$ of this total input data (i.e., the amount of data
to be processed by request~$r$ is $\zeta_r \bar{u}_{k,s}$).
Furthermore, the computing load of each function $l{\in}F_r$ of request~$r$
can be computed based on the computing load per input data bit $c^l_{k,s}$.


\subsubsection{Local Computation Models}

Let $f^{r,l}_{k,s}$ be the computing resource (CPU clock speed) allocated by MU~$k$ at BS~$s$ to execute
function $l$ of request $r$ locally at the MU. We assume that $f^{r,l}_{k,s}$ must be chosen in the range $(0,\bar{F}_{k,s}]$ 
where $\bar{F}_{k,s}$ denote the MU's maximum CPU clock speed (i.e., computing capacity).
Then, the processing delay and energy consumption for local execution of request $r$ of MU $(k, s)$
can be expressed, respectively as
\begin{equation} 
\hspace{1mm}\small{\Delta^{\mathrm{\textbf{loc}},r}_{k,s}} {=} \small{\sum \limits_{\mathclap{l \in F_{r}}}} \small{{\zeta_r \xi_{k,s}^{r,l} c^l_{k,s} \bar{u}_{k,s}}{/}{f^{r,l}_{k,s}}};  \small{e^{\mathrm{\textbf{loc}},r}_{k,s}}{=} \small{\sum \limits_{\mathclap{l \in F_{r}}}} \zeta_r \xi_{k,s}^{r,l} \bar{u}_{k,s} \kappa_{k,s} {f^{r,l}_{k,s}}^2 \hspace{-1.5mm}\label{loc_e_t}
\end{equation}


\subsubsection{Computation Offloading Models}

\begin{figure}
\centering
\includegraphics[width=0.8\columnwidth]{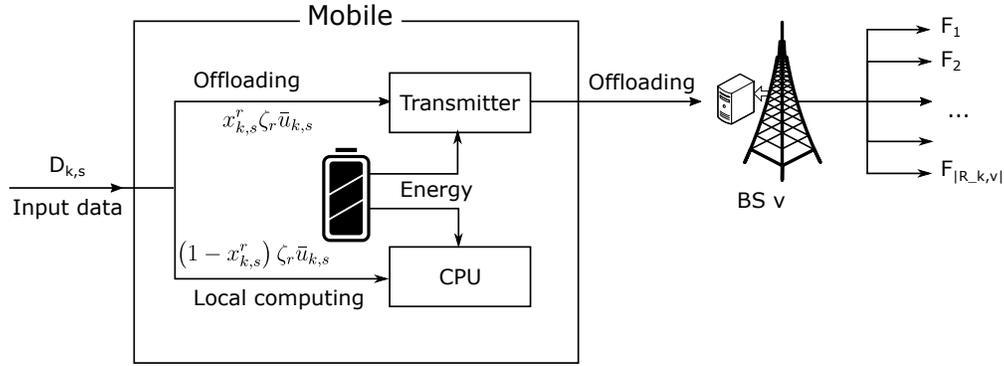}
\caption{Mobile computation offloading model}
\label{fig-Offl-mdl}
\vspace{-5mm}
\end{figure}

The mobile computation offloading scheme is illustrated in Figure \ref{fig-Offl-mdl}. 
The data of each offloaded request $r \in R_{k,s}$ must be first transmitted to the associated BS $s$. The
request can be either processed by the CoS at this BS or sent out to its neighboring CoSs for execution. 
The total execution delay $T_{k,s}$ of the application of MU $(k,s)$ is defined as the maximum execution delay of
individual requests either at the MU or at remote CoSs via offloading. And this execution delay must be
constrained by the maximum allowable delay $\bar{T}_{k,s}$:
\begin{equation}
T_{k,s} = \max \limits_{r \in R_{k,s}} \left(\Delta^{\mathrm{\textbf{loc}},r}_{k,s}, \Delta^{\textbf{ofl},r}_{k,s} \right) \leq \bar{T}_{k,s}
\end{equation}
where $\Delta^{\mathrm{\textbf{loc}},r}_{k,s}$ and $\Delta^{\textbf{ofl},r}_{k,s}$ denote the execution delay
of request $r$ if done locally at the MU or at remote CoSs via offloading, respectively. We show how to calculate
$\Delta^{\textbf{ofl},r}_{k,s}$ in the following.

To enable the offloading of any particular request $r$ of MU $(k, s)$,
the involved data $\zeta_r \bar{u}_{k,s}$ must be transmitted in the uplink direction
from MU $k$ to BS $s$. Recall that we consider the multi-cell Massive-MIMO wireless system 
where each MU has a single antenna and each BS is equipped with  $M_s$ antennas where $M_s \gg | \mathcal{K}_s |$.  
We assume that the same transmit power $p$ is used by each MU to transmit the training data (to estimate the channel state information)
and application data (to support the offloading). The achieved signal to interference ratio (SIR) of the uplink transmission
 from MU $k$ in the cell $s$ can be expressed as $\text{SIR}_{sk} {=}  {\beta^2_{sks}}/{\sum_{q \neq s} \beta^2_{qks}}$ \cite{marzetta2010noncooperative}, where $\beta_{qsk}{=}1 {/} r^{\gamma}_{qks}$ represents
the large-scale channel coefficient capturing the path-loss effect,
$r_{qks}$ is the distance between the co-channel MU in cell $q$ of MU $k$ and the BS $s$, and $\gamma$ is the 
path-loss exponent. 
Then, the corresponding achievable rate can be written as $r_{s,k} {=} W_s \log_2 \left( 1{+} \text{SIR}_{sk}\right)$.

For request $r$, the total execution time is the sum
of the uplink communication delay $t^{r}_{k,s}$, the backhaul transfer delay $\Delta^{r,l,fcm}_{\bar{u}_{k,s}}$ and processing delay $\Delta^{r,l,fcp}_{\bar{u}_{k,s}}$ of all network functions
in the SFC of request $r$. Thus, we have
\begin{equation}
\Delta^{\mathrm{\textbf{ofl}},r}_{k,s} = t^{r}_{k,s} + \sum \limits_{l \in F_r} \left(\Delta^{r,l,fcm}_{\bar{u}_{k,s}} + \Delta^{r,l,fcp}_{\bar{u}_{k,s}} \right)
\end{equation}
where $t^{r}_{k,s}{=} {\zeta_r \bar{u}_{k,s}}/{r_{s,k}}$. The energy consumption required to transmit the involved data for offloading can be calculated $e^{\mathrm{\textbf{tx}},r}_{k,s} {=} p{*}t^r_{k,s}$. Detailed descriptions on how these delay components can be calculated are given in the following.

\subsection{Offloading Parameters and SFC Placement Constraints}

Different network functions associated with request $r$ of MU $(k, s)$ can be processed at CoS associated with BS $s$ or routed to neighboring CoSs with larger computation resource for processing.
The network operator must make decisions on request offloading as well as placement and execution of different functions of each request. 
Toward this end, we introduce three optimization variable sets. The first set of variables $x^r_{k,s}$ 
represents the binary offloading decisions where if request $r$ of MU $(k, s)$ is processed locally at this MU then $x^r_{k,s}{=}0$; otherwise, we have $x^r_{k,s}{=}1$ if the request is  offloaded to remote CoSs. The second variable set $A^{r,l}_{k,s,m} {\in} \{0,1\}$ indicates the SFC placement where if function $l {\in} F_r$ of request $r {\in} R_{k,s}$ of MU $(k,s)$ is placed at CoS $m$, we have 
$A^{r,l}_{k,s,m}{=}1$; otherwise, we have $A^{r,l}_{k,s,m}{=}0$.
The last variable set represents the computing resource allocation (in CPU clock speed)
where $f^{r,l}_{k,s},f^{r,l}_{k,s,m}$ denote the CPU clock speeds assigned to serve function $l$ of request $r$
locally at MU $(k,s)$ or remotely at CoS $m$, respectively.



The function placement needs to satisfy several constraints:

\vspace{0.9mm}
\subsubsection{Function placement constraints}
Each function should be placed at exactly one CoS:

$\sum_{m \in \mathcal{V}} x^r_{k,s} A^{r,l}_{k,s,m}{=}x^r_{k,s},\;\; \forall l {\in} F_r,\;\; \forall r{\in} R_{k,s}, \forall k {\in} \mathcal{K}_v, v {\in} \mathcal{V}$

The total computation load routed to CoS $m$ must not exceed its computing capacity:

\hspace{-2mm}$ \sum_{s \in \mathcal{S}} \sum_{k \in \mathcal{K}_s} \sum_{r \in R_{k,s}} \sum_{l \in F_r} x^r_{k,s} A^{r,l}_{k,s,m} f^{r,l}_{k,s,m}  {\leq} \bar{F}_m, \forall m {\in} \mathcal{V}.$

\vspace{0.9mm}
\subsubsection{Routing path constraints}
the functions $l$ and $l{+}1$ associated with request $r$ of MU $(k, s)$ must be 
placed at the same CoS or at two inter-connected CoSs. Therefore, 
we have the following constraints for $A^{r,l}_{k,s,v}{\times}A^{r,l+1}_{k,s,m}$:
\hspace{-2mm}$\sum_{m \in  \mathcal{V} } \sum_{p \in \mathcal{V} } x^r_{k,s}  \mathbbm{e}_{mp} ( A^{r,l}_{k,s,m} {\times} A^{r,l+1}_{k,s,p}) {=} x^r_{k,s}, \forall l {\in} F_r, \forall r {\in} R_{k,s}$.

By applying the offloading parameters, we represent  the backhaul transfer delay for the involved data between CoSs $m$ and $p$ can be expressed as:
\begin{equation}
\label{eq:fcm-delay}
\begin{split}
&\Delta^{r,l,fcm}_{\bar{u}_{k,s}} = \sum \limits_{m \in \mathcal{V}} \sum \limits_{p \in \mathcal{V}} (A^{r,l}_{k,s,m} \times A^{r,l+1}_{k,s,p})  \delta^{cm}_{mp}
\end{split}
\end{equation}

The server computation time of function $l$ is a function of the allocated computing resource and it can be expressed as:
\begin{equation}
\label{eq:fcp-delay}
\begin{split}
&\Delta^{r,l,fcp}_{\bar{u}_{k,s,m}} = \sum \limits_{m \in \mathcal{V}} {(A^{r,l}_{k,s,m} \zeta_r \xi_{k,s}^{r,l} \bar{u}_{k,s}c^l_{k,s})}/{f^{r,l}_{k,s,m}}
\end{split}
\end{equation}

\subsection{Problem Formulation}

Our design aims to minimize: 1)
normalized mobile energy consumption, and 2) normalized computing cost. 
Toward this end, we will optimize a single objective function which is the weighted sum of these two optimization
metrics of interest. 
The normalized mobile energy consumption is equal to ratio between the energy consumption 
and the total energy pool $\mathbb{E}_{k,s}$ where the energy consumption is equal to either
the  local computing energy or communication energy depending on the offloading decision.

The offloading/computing cost is calculated based on the computing price per time unit
at the processing speed $f$ which is expressed as $P^{\mathrm{\textbf{cp}}}(f){=}e^{-\eta}(e^{f}{-}1)\vartheta$ \cite{du2017contract}.
Then, the computing cost required to process function $l$ of request $r$ of MU  $(k, s)$ at CoS $m$ can be expressed as 
$C(f^{r,l}_{k,s,m}){=}P^{\mathrm{\textbf{cp}}}(f^{r,l}_{k,s,m}) * T^{\mathrm{\textbf{cp}}}$ 
where the required computing time of the corresponding function is $T^{\mathrm{\textbf{cp}}}{=}{\zeta_r \xi_{k,s}^{r,l} \bar{u}_{k,s}c^l_{k,s}}/{f^{r,l}_{k,s,m}}$. 
The coefficients $\eta$ and $\vartheta$ can vary with cloud platforms.
Suppose the available budget to cover the computing expenses is ${\mathbb{C}_{k,s}}$. 

To maintain the total system utility, an amount of budget is given and is granted equally for all MUs. The normalized energy consumption and VNF placement cost is accounted for the ratio of used computation and is scaled to normally common budget of energy $\mathbb{E}_{k,s}$  and computing cost ${\mathbb{C}_{k,s}}$. Then, we define the normalized cost of each MU  $(k, s)$ as:
\begin{align}
\hspace{-4mm} Z_{k,s}&{=} Z^{\mathrm{\textbf{loc}}}_{k,s} {+} Z^{\mathrm{\textbf{off}}}_{k,s} {=} \sum \limits_{r \in R_{k,s}} [\left( ( 1- x^r_{k,s} \right)  {e^{\mathrm{\textbf{loc}},r}_{k,s}/\mathbb{E}_{k,s}} {+} x^r_{k,s} ( \theta^{\mathrm{\textbf{tx}}} {e^{\mathrm{\textbf{tx}},r}_{k,s}{/}\mathbb{E}_{k,s}} {+} \theta^{\mathrm{\textbf{\textbf{cp}}}} C^{r,l}_{k,s,m}(f^{r,l}_{k,s,m}){/}{\mathbb{C}_{k,s}})]
\end{align}
where $\theta^{\mathrm{\textbf{\textbf{tx}}}}$ and $\theta^{\mathrm{\textbf{\textbf{cp}}}}$ are weighting parameters capturing the
importance of energy consumption and computing cost, respectively where $\theta^{\mathrm{\textbf{\textbf{tx}}}} + \theta^{\mathrm{\textbf{\textbf{cp}}}} = 1$.

The considered  \textsc{J}oint 
\textsc{C}omputation \textsc{O}ffloading and \textsc{R}esource \textsl{A}llocation (JCORA) problem  can be 
formulated as follows:

\vspace{0.1cm}
\noindent 
\textbf{Problem ($\textbf{JCORA}$):}
\begin{subequations} \label{p1}
\begin{eqnarray} 
\hspace{-8mm} &  \min\limits_{\Omega_1} & \sum \limits_{s \in \mathcal{S}}  \sum \limits_{k \in \mathcal{K}_s}  Z_{k,s} \nonumber\\[-0 pt]
\hspace{-8mm}&\text{s.t } &  \sum \limits_{r \in R_{k,s}} \left( \left( 1- x^r_{k,s} \right) \sum \limits_{l \in F_{r}} {f^{r,l}_{k,s}} \right) \leq \bar{F}_{k,s}, \forall (k,s) \label{JCORAcnstLocalCompMax}\\[-0 pt]
\hspace{-8mm}&&  \max \limits_{r \in R_{k,s}} \left( \left( 1- x^r_{k,s} \right) \Delta^{\mathrm{loc},r}_{k,s},  x^r_{k,s} \left( {\zeta_r \bar{u}_{k,s}}/{r_{s,k}} + \sum \limits_{l \in F_r} \left(\Delta^{r,l,fcm}_{\bar{u}_{k,s}} + \Delta^{r,l,fcp}_{\bar{u}_{k,s}} \right) \right) \right) \leq \bar{T}_{k,s},\forall (k,v) \hspace{4mm} \label{JCORAcnstDelay}\\[-0 pt]
\hspace{-9mm}&&\sum \limits_{m \in V} x^r_{k,s} A^{r,l}_{k,s,m} = x^r_{k,s}, \forall l \in F_r, \forall r \in R_{k,s},  \forall (k,v) 
\label{JCORAcnstPlcmntBin}\\[-0 pt]
\hspace{-15mm}&& \sum \limits_{s \in \mathcal{S}} \sum \limits_{k \in \mathcal{K}_s} \sum \limits_{r \in R_{k,s}} \sum \limits_{l \in F_r} x^r_{k,s} A^{r,l}_{k,s,m} f^{r,l}_{k,s,m}  \leq \bar{F}_m, \forall m \;\;\;\;\;\;\;\;\;\label{JCORAcnstEdgeCompMax}\\[-0 pt]
\hspace{-8mm}&& \sum \limits_{m \in  \mathcal{V} } \sum \limits_{p \in  \mathcal{V} } x^r_{k,s} \mathbbm{e}_{mp} \left( A^{r,l}_{k,s,m} \times A^{r,l+1}_{k,s,p} \right) = x^r_{k,s}, \forall l,r \label{JCORAcnstEdgeBound1}\\[-0 pt]
\hspace{-8mm}&& A^{r,l}_{k,s,m} = 0 , \forall l \in \mathfrak{F}_m, \forall m \in \mathcal{V} \setminus \mathcal{S} \\[-0 pt]
\hspace{-8mm}&& A^{r,l}_{k,s,m} \in \{0,1\} \label{JCORAvarA}\\[-0 pt]
\hspace{-8mm}&& x^r_{k,s} \in \{0,1\}\label{JCORAvarX}
\end{eqnarray}
\end{subequations}
where the set of optimization variables is defined as $\Omega_1 {=} \{x^r_{k,s},f^{r,l}_{k,s},A^{r,l}_{k,s,m},f^{r,l}_{k,s,m} \}$.

\section{Proposed Algorithm}

We describe our proposed algorithm to solve problem (JCORA) in this section.
Problem (JCORA) is difficult to solve because it is a mixed integer and non-linear optimization problem.
Specifically, there are two set of variables concerning the local resource allocation $\{f^{r,l}_{k,s}\}$ and the 
SFC placement and resource allocation at the CoSs $\{A^{r,l}_{k,s,m},f^{r,l}_{k,s,m}\}$. 
To solve problem (JCORA), we employ the decomposition approach where we optimize different sets of variables
separately by tackling the corresponding sub-problems in the iterative manner.
The proposed iterative algorithm is given in Algorithm~\ref{alg:InterativeProc}.
This algorithm has an initialization step in which we try to execute as many
requests locally at MUs as possible while using all local computing resource (step 0).
After initialization, we know the set of requests executed locally (called local request set) and the set of
requests offloaded to remote servers (called offloading request set). We then optimize the function chain
placement for all offloaded requests and computing resource allocation for them (step 1).
To further improve the performance, we iteratively update the offloading decisions by
moving more requests from MUs to the remote CoSs (step 2).
We describe these steps in more details in the following.

\subsection{Step 0: Initialization at MUs} 
For initialization, we attempt to minimize the total local computation energy
by solving the following problem:

\vspace{0.1cm}
\noindent
\textbf{Problem\; ($\textbf{JPL}$):}
\begin{subequations} \label{p2}
\begin{eqnarray} 
&\min\limits_{\tiny{x^r_{k,s},{f^{r,l}_{k,s}}}} &\sum \limits_{r \in R_{k,s}} \sum \limits_{l \in F_{r}} (1 - x^r_{k,s}) \zeta_r \xi_{k,s}^{r,l} \bar{u}_{k,s} \kappa_{k,s} ({f^{r,l}_{k,s}})^2/\mathbb{E}_{k,s}  \nonumber\\[-0 pt]
&\text{s.t }   &(\ref{JCORAcnstLocalCompMax}),(\ref{JCORAcnstDelay}). \nonumber
\end{eqnarray}
\end{subequations}
We solve this problem by first tackling the local computation allocation for all requests by assuming that
the local computing capacities at all MUs are very large. Then, we use this computation allocation result
to determine the local request set and offloading request set. These two sub-steps are as follows.


\begin{algorithm}[!t]
\footnotesize
\caption{\textsc{Iterative Algorithm To Solve Problem (JCORA)}}
\label{alg:InterativeProc}
\algsetup{indent=1.5em}
\begin{algorithmic}[1]
\STATE \textbf{Step 0} Compute local computation resource allocation $f^{r,l}_{k,s}$ as given in (\ref{jpl_f_opt})  by 
solving problem (\textbf{JPL1}) and initialize offloading decision $x^r_{k,s}$ by solving problem (\textbf{JPL2}) 
\STATE \textbf{Step 1} Determine function placement $A^{r,l}_{k,s,m}$ and computation allocation $f^{r,l}_{k,s,m}$ by solving problem (JPE) as described in Algorithm \ref{alg:JPE_Decomposing}.
\STATE \textbf{Step 2} Find $\{\hat{s},\hat{k},\hat{r}\}=\arg\max \limits_{s,k,r}{\left( 1- x^r_{k,s} \right) \Delta Z(k,s,r)}$
\IF {$ \{\hat{s},\hat{k},\hat{r}\} \neq \emptyset$ and $\Delta Z(\hat{s},\hat{k},\hat{r}) >$ 0 }
\STATE Update offloading decision $x^{\hat{r}}_{\hat{k}.\hat{s}}=1$
\STATE Goto \textbf{Step 1}
\ELSE 
\STATE Terminate and set \textbf{Output}$=\{x^r_{k,s},f^{r,l}_{k,s},A^{r,l}_{k,s,m},f^{r,l}_{k,s,m}\}$
\ENDIF
\end{algorithmic}
\end{algorithm}

\vspace{0.5cm}
\noindent
\textbf{Sub-step 1 - Local computation resource allocation}

To determine local computation allocation, we solve problem ($\textbf{JPL1}$)
with the same objective with problem ($\textbf{JPL}$) assuming that $x^r_{k,s} = 0, \; \forall k, s, r$
considering only constraints (\ref{JCORAcnstDelay}). 
The Lagrangian for problem ($\textbf{JPL1}$)  \cite{boyd2004convex} can be written as
\begin{align}
\small{L_1 {=}} \small{\sum \limits_{{r{\in}{R_{k,s}}}}}\small{\sum \limits_{{l \in F_{r}}}} \small{\sdfrac{\zeta_r \xi_l \bar{u}_{k,s} \kappa_{k,s} ({f^{r,l}_{k,s}})^2}{\mathbb{E}_{k,s}}{+}\small{\lambda_{r}} \left[\sdfrac{ \sum \limits_{\mathclap{l \in F_{r}}}  c^l_{k,s} \zeta_r 
\xi_{k,s}^{r,l} \bar{u}_{k,s}}{f^{r,l}_{k,s}}{-}\bar{T}_{k,s}\right]}\nonumber
\end{align}
Taking the derivative of the Lagrangian  w.r.t $f^{r,l}_{k,s}$, we have:
\begin{align}
\sdfrac{\partial L_1}{\partial f^{r,l}_{k,s}} = \sdfrac{2 \zeta_r \xi_{k,s}^{r,l} \bar{u}_{k,s} \kappa_{k,s} {f^{r,l}_{k,s}}}{\mathbb{E}_{k,s}} - \lambda_{r}  \sdfrac{c^l_{k,s} \zeta_r \xi_{k,s}^{r,l} \bar{u}_{k,s}}{{f^{r,l}_{k,s}}^2} \label{jpl_f_trans}
\end{align}

Setting this derivative to 0 yields the estimation of$f^{r,l}_{k,s}{=}{({\lambda_{r}{\mathbb{E}_{k,s}}{c^l_{k,s}}}/{2\kappa_{k,s}})}^{1/3}$.
It can be verified that the objective function of problem (JPL1)  is non-decreasing with the allocated  computing resource, thus,
at the optimal ${f^{*_{r,l}}_{k,s}}$, the equality condition for (\ref{JCORAcnstDelay}) holds; thus, $T_{k,s}{=}\bar{T}_{k,s}$.  From this condition, we can obtain the allocated computing resource as follows \cite{lyu2016multiuser}:
\begin{equation} 
f^{r,l}_{k,s}=({\zeta_r \xi_{k,s}^{r,l} \bar{u}_{k,s} {\mathbb{E}_{k,s}}\sqrt[3]{c^{l^{4}}_{k,s}}\sum_{l \in F_{r}} \sqrt[3]{{c^{l^{-1}}_{k,s}}}})/{\bar{T}_{k,s}}\label{jpl_f_opt}
\end{equation}

\noindent
\textbf{Sub-step 2 - Determination of local/offloading request sets}

Using the computation allocation results for (\refeq{jpl_f_opt}) in problem ($\textbf{JPL1}$),
we arrive at the following problem:

\noindent
\textbf{Problem\; (JPL2):}
\begin{subequations} \label{p2}
\begin{eqnarray} 
&\min\limits_{\tiny{x^r_{k,s}}} \sum \limits_{r \in R_{k,s}} \sum \limits_{l \in F_{r}} (1 - x^r_{k,s}) \zeta_r \xi_{k,s}^{r,l} \bar{u}_{k,s} \kappa_{k,s} ({f^{r,l}_{k,s}})^2/\mathbb{E}_{k,s}  \nonumber\\[-0 pt]
&\text{s.t }  \sum \limits_{r \in R_{k,s}} \left( 1- x^r_{k,s} \right) \sum \limits_{l \in F_{r}} {f^{r,l}_{k,s}}  \leq \bar{F}_{k,s}, \forall (k,s)  \nonumber
\end{eqnarray}
\end{subequations}

This is indeed a knapsack problem which determines the requests to be executed locally at each MU (i.e., requests with $x^r_{k,s}{=}0$)
where the size of each item/request is the total computation resource required by its functions, i.e., 
 $\textit{item\_size}^{r}_{k,s}{=}\sum_{l \in F_r} f^{r,l}_{k,s}$. The knapsack problem can be efficiently solved via ILP solver \cite{optimization2014inc} which will try to pack as many items (requests) as possible to fill up the bin size $\bar{F}_{k,s}$ and it stops at the split point. The remaining items/requests will be offloaded to remote CoSs.

\subsection{Step 1: Function chain placement and computation resource allocation at remote CoSs}

After step 0, we obtain the set of offloaded requests of each
MU $(k, s)$ which is denoted by $R^{\mathrm{\textbf{off}}}_{k,s}$.
The function chain placement and computation resource allocation
for all functions of these offloaded requests can be determined
by solving the following problem:

\vspace{0.1cm}
\noindent
\textbf{Problem\; (JPE):} 
\begin{subequations} \label{p3}
\begin{eqnarray} 
\hspace{-8mm}\small{\min_{\mathclap{\Omega_2}}} &&\hspace{-3mm} \sum_{\mathclap{s \in \mathcal{S}}}  \sum_{\mathclap{k \in \mathcal{K}_s}} \small{\sum_{ r \in R^{\mathrm{\textbf{off}}_{k,s}}} \left[ \small{\theta^{\mathrm{\textbf{tx}}}}  \sdfrac{\bar{p}\zeta_r \xi_{k,s}^{r,l} \bar{u}_{k,s}}{\mathbb{E}_{k,s}r_{k,s}}{+}\small{\theta^{\mathrm{\textbf{\textbf{cp}}}}} \small{\sum_{\mathclap{l \in F_r}} \sum_{\mathclap{m \in V}}} \sdfrac{A^{r,l}_{k,s,m}C(f^{r,l}_{k,s,m})}{\mathbb{C}_{k,s}} \right]} \nonumber\\[-0 pt]
\hspace{-8mm}\text{s.t } &&\hspace{-3mm} \max \limits_{r \in R_{k,s}} \left( \sdfrac{\zeta_r \bar{u}_{k,s}}{r_{s,k}} + \sum \limits_{l \in F_r} \left[\sum \limits_{m \in \mathcal{V}} \sum \limits_{p \in \mathcal{V}} \left(A^{r,l}_{k,s,m}{\times} A^{r,l+1}_{k,s,p}\right)  \delta^{\mathrm{\textbf{tx}}}_{mp}  {+}  \sum_{\mathclap{m \in V}} \sdfrac{A^{r,l}_{k,s,m} \zeta_r \xi_{k,s}^{r,l} \bar{u}_{k,s}c^l_{k,s}}{f^{r,l}_{k,s,m}} \right]  \right) \leq \bar{T}_{k,s},\forall (k,s) \hspace{10mm}\label{JPEcnstDelay}\\[-0 pt]
\hspace{-3mm}&&\hspace{-3mm}\text{and}\; (\ref{JCORAcnstPlcmntBin}){-}(\ref{JCORAvarA})\nonumber
\end{eqnarray}
\end{subequations}
where $\Omega_2{=}\{A^{r,l}_{k,s,m},f^{r,l}_{k,s,m}\}$ and $R^{\mathrm{\textbf{off}}}_{k,s}$ denotes the set of offloaded requests of
MU $(k, s)$.

This problem is a mixed integer optimization problem and still hard to solve. To tackle the problem, we employ Bender's 
decomposition approach that separates the original problem into a \textit{slave} problem for computation
resource optimization and a \textit{master} problem for function placement 
optimization. The proposed algorithm is summarized in Algorithm~2. 
Detailed descriptions of the master and slave problems are given in the following.


\subsubsection{\textbf{Master} problem to optimize function placement $A^{r,l}_{k,s,m}$}

\noindent
\textbf{Problem \; ($\textbf{JPE}_{\textbf{M}}$):}
\begin{subequations} \label{p2_2fixtimeslot}
\begin{eqnarray}
\hspace{-3mm} &\min\limits_{A^{r,l}_{k,s,m}} & \sum_{{s \in \mathcal{S}}}  \sum_{k \in \mathcal{K}_s} \small{\sum_{r \in R^{\mathrm{\textbf{off}}}_{k,s}}}
\sum_{\mathclap{l \in F_r}} \sum_{\mathclap{m \in V}} {A^{r,l}_{k,s,m}C(f^{r,l}_{k,s,m})}/{\mathbb{C}_{k,s}}  \label{JPEM} \nonumber\\[-0 pt]
\hspace{-3mm}&\text{s.t } & (\ref{JPEcnstDelay}), (\ref{JCORAcnstPlcmntBin}){-}(\ref{JCORAvarA}) \nonumber
\end{eqnarray}
\end{subequations}
Similar to step 0, to solve this problem, we estimate the computation resource allocation
for all functions of offloaded requests in the first sub-step; then, using this result,
we determine the service function placement solution in the second sub-step.


In the \textbf{first sub-step}, we solve a related problem of Problem ($\textbf{JPE}_{\textbf{M}}$) where it has the 
same objective $\min  \sum_{{s \in \mathcal{S}}} \sum_{k \in \mathcal{K}_s}\penalty 0  \sum_{r \in R^{\mathrm{\textbf{off}}}_{k,s}}
\sum_{l \in F_r}  C(\tilde{f}^{r,l}_{k,s})/{\mathbb{C}_{k,s}}$
subject to the delay constraints (\ref{JPEcnstDelay}). Here, we assume that the maximum computing resource
at each CoS is sufficiently large; therefore, the computation resource allocation is performed
to achieve the minimum computation cost while simply maintaining the delay constraints. As a result, the considered 
computation resource allocation variables $\tilde{f}^{r,l}_{k,s}$ do not depend on the CoS index $m$.

We solve this problem by defining the Lagrangian and solve
the Karush-Kuhn-Tucker optimality conditions \cite{boyd2004convex}.
After several manipulations, we can derive the following computation resource allocation policy:
\begin{equation}
\tilde{f}^{r,l}_{k,s} (\mu^{r}_{k,s}) = W_n \left( \frac{\mathbb{C}_{k,s}e^{-\eta}\vartheta \mu^{r}_{k,s}-1}{e}\right) + 1
\label{jpe_f_opt}
\end{equation}
where $W_n(\cdot)$ is the Lambert function \cite{corless1996lambertw} and $\mu^{r}_{k,s}$ can be obtained by solving the following equation:
\begin{equation}
\sum \limits_{l \in F_r} \sdfrac{\zeta_r \xi_{k,s}^{r,l} \bar{u}_{k,s}c^l_{k,s}}{W_n \left( \sdfrac{\mathbb{C}_{k,s} e^{-\eta}\vartheta \mu^{r}_{k,s}-1}{e} \right) + 1} = \bar{T}_{k,s} -  \sdfrac{\zeta_r \bar{u}_{k,s}}{r_{s,k}} - \sum \limits_{l \in F_r}  \delta^{\mathrm{\textbf{tx}}}_{mp}. \nonumber
\end{equation}
Hence, the root of this equation can be determined by using a numerical 
searching method.


In the \textbf{second sub-step}, we perform function placements by solving another related 
problem with Problem ($\textbf{JPE}_{\textbf{M}}$) where it has the same 
objective with ($\textbf{JPE}_{\textbf{M}}$) but with only constraints (\ref{JCORAcnstPlcmntBin}){-}(\ref{JCORAvarA}).
The computation allocation solution obtained in the first sub-step is used to
estimate the consumed computation resources during the function placements.
This problem is more complicated than the multi-knapsack problem
due to the additional backhaul topology constraints (\ref{JCORAcnstEdgeBound1}).

To solve this problem, we propose a greedy function placement algorithm which is described in Algorithm~2.
This algorithm has two phases. In phase one, we attempt to place functions of offloaded requests 
at the corresponding local CoSs of BSs. This is done by solving the knapsack problem with the local maximum 
computation constraint. 
In phase two, we perform placements for the remaining (un-placed) network functions denoted as $F_{\text{un}}$, which have not been placed
in phase one. To efficiently utilize CoSs' computing resource leveraging the load balancing, it is desired to
place more functions to CoSs with larger available computing resource and being connected with a
smaller number of neighboring CoSs. 

After phase one, let $F^{\text{free}}_m$ denote the remaining
computing resource in CPU clock speed of CoS $m$, which is equal to $\bar{F}_m$ minus the total
estimated computing resource $\tilde{f}^{r,l}_{k,s}$ of all functions $l$ placed at CoS $m$ in phase one 
where $\tilde{f}^{r,l}_{k,s}$ is given in (\ref{jpe_f_opt}). We define the ranking metric for each
CoS $m$ as $\text{MP}_m=F^{\text{free}}_m / \vert \mathcal{L}^I_m \vert$ where $\mathcal{L}^I_m$ is the in-neighbor CoS set
 of CoS $m$. We then rank CoSs in the descending order of $\text{MP}_m$ and let 
$\mathcal{\tilde{V}}$ denote the corresponding ordered set of CoSs. Then, for each
CoS $m$ in the ordered set of CoSs $\mathcal{\tilde{V}}$, we perform function placements
by solving the corresponding knapsack problem whose objective is to minimize the
total computation cost subject to the constraint on the (remaining) computing capacity.
After performing function placements for all CoSs in $\mathcal{\tilde{V}}$, we obtain the function placement solution
(i.e., ${A}^{r,l}_{k,s,m}$).

\begin{algorithm}[ht]
\footnotesize
\caption{\textsc{Greedy Topology Decomposition Algorithm (GTDA) To Solve Problem (JPE)}}
\label{alg:JPE_Decomposing}
\algsetup{indent=1.5em}
\begin{algorithmic}[1]
\STATE \textbf{Step 1} \label{alg2:stp0} Determine estimated computation allocation as given in (\ref{jpe_f_opt})
\STATE \textbf{Step 2.1} \label{alg2:stp1} Perform function placements for each edge server $m \in \mathcal{S} $


\STATE \textbf{Step 2.2} \label{alg2:stp3} Perform function placements for remaining functions as follows.


\FOR {each CoS $m \in \mathcal{\tilde{V}}$}
\STATE  \label{alg2:stp3b} Solve the knapsack problem at CoS $m$ to determine
the functions from $F_{\text{un}}$ to be placed at this CoS
\ENDFOR
\STATE \textbf{Step 3} \label{alg2:stp6} Solve problem $({JPE}_{S})$ to obtain final computing
resource allocation solution $f^{r,l}_{k,s,m}$ by using CVX solver
\end{algorithmic}
\end{algorithm}

\subsubsection{\textbf{Slave} problem to optimize computation resource allocation $f^{r,l}_{k,s,m}$}
For given $\bar{A}^{r,l}_{k,s,m}$, we introduce slack variable $y^{r,l}_{k,s,m} = \delta^{\mathrm{\textbf{tx}}}  + \zeta_r \xi_{k,s}^{r,l} \bar{u}_{k,s}c^l_{k,s}/{f^{r,l}_{k,s,m}}$. Then, the \textit{slave} problem that optimizes
the computation resource allocation can be stated as:

\noindent
\textbf{Problem\; ($\textbf{JPE}_{\textbf{S}}$): }
\begin{subequations} \label{p2_2GA}
\begin{eqnarray}
\hspace{-7mm}&\min\limits_{\mathclap{y^{r,l}_{k,s,m}}} & \sum \limits_{s \in \mathcal{S}}  \sum \limits_{k \in \mathcal{K}_s} \sum \limits_{r \in R^{\mathrm{\textbf{off}}}_{k,s}} \beta_{k,s} \left[ e^{-\eta}\left(e^{f^{r,l}_{k,s,m}}-1\right)({y^{r,l}_{k,s,m} - \delta^{\mathrm{\textbf{tx}}}}) \right]\nonumber\\[-0 pt]
\hspace{-7mm}&\text{s.t } & \sum \limits_{l \in F_r} y^{r,l}_{k,s.m} = \bar{T}_{k,s} - \frac{\bar{u}_{k,s}}{r_{k,s}}, \forall (k,s), \forall r \in R_{k,s} \label{J3ScnstDelay} \\[-0 pt]
\hspace{-7mm}&& \small{\sum_{\mathclap{s \in \mathcal{S}}} \sum_{\mathclap{k \in \mathcal{K}_s}} \sum_{r \in R_{k,s}} \sum_{\mathclap{l \in F_r}}} (y^{r,l}_{k,s,m}{-}\delta^{\mathrm{\textbf{tx}}})^{\small{{-}1}}\zeta_r \xi_{k,s}^{r,l} \bar{u}_{k,s}c^l_{k,s}{\leq}\bar{F}_m \label{J3ScnstEdgeCompMax}\\[-0 pt]
\hspace{-7mm}&& y^{r,l}_{k,s.m} - \frac{1}{f^{r,l}_{k,s,m}} \leq 0
\end{eqnarray}
\end{subequations}

Problem ($\textbf{JPE}_{\textbf{S}}$) is a convex optimization problem due to its affine equality constraints, convex objective function
 and convex equality constraint function. Thus, it can be solved efficiently to obtain the optimal values of $y^{{opt}_{r,l}}_{k,s,m}$ 
and $f^{{opt}_{r,l}}_{k,s,m}$.

\subsection{Step 2: Update offloading decisions} 

To update the offloading decisions, we define the following cost improvement factor $\Delta Z$:
\begin{equation}
\Delta Z(k,s,r){=}\sdfrac{e^{\mathrm{\textbf{loc}},r}_{k,s}}{\mathbb{E}_{k,s}}{-}
  ( \theta^{\mathrm{\textbf{tx}}} \sdfrac{e^{\mathrm{\textbf{tx}},r}_{k,s}}{\mathbb{E}_{k,s}} + \theta^{\mathrm{\textbf{\textbf{cp}}}}  \sdfrac{C(f^{r,l}_{k,s,m})}{\mathbb{C}_{k,s}})
\end{equation}
which quantifies the cost reduction if we offload request $r$ to the CoSs.

Specifically, in step 2 of the proposed algorithm, we iteratively and greedily
find one request $r$ with positive and maximum cost reduction $\Delta Z(k,s,r)$ 
where this request is currently executed locally and we force this request to be offloaded to remote CoSs. 

\section{Numerical Results}

We consider a simple 4-cell network where the distance between two nearest BSs is $1600 \, m$ as illustrated in Fig.~\ref{fig-MultiFog}. 
In each cell, we randomly place 8 MUs 
 so that the distance from the BS to its MUs is in the range $[100 m,800 m]$.
The channel gains are generated by considering path-loss exponent $\gamma=3.8$.
In the simulation, we choose $\beta_{k,s}$ equal to $1$ for all MUs, $W_s=300$~kHz 
and $\delta^{\mathrm{\textbf{tx}}}_{mp} = 10$~ms. 
Each MU needs to execute an application with data size of 800 kbits ($u_{k,s}=0.8$Mbit) within the maximum delay of $800$~ms 
($\bar{T}_{k,s}=0.8s, \forall (k,s)$) where each application is assumed to be split
into 5 requests ($\vert R_{k,s}\vert=5, \forall (k,s)$). 

The maximum computing capacity for each MU ($\bar{F}_{k,s}$) is randomly selected from the set $\{0.5, 0.4,..., 0.8\}$~GHz and the local computing energy per CPU cycle is $\kappa_{k,s}=10^{-26}$~J/CPU cycle. Each data bit is assumed to consume
 $c^l_{k,s} \in \left[ 200,500 \right]$ CPU cycle/bit. Finally, the capacity of four servers are chosen as $\{1.7, 3.6, 3.8, 4.5\}$ GHz. 
The energy and computing budgets of each MU are allocated as $\mathbb{E}_{k,s}=100$mW and $\mathbb{C}_{k,s}=0.035\$$ 
which are set based on the cost of Amazon AWS and IBM clouds, which yields the cost of $\eta=1$ and $\vartheta=2.5.10^{-12}$\$ per one CPU clock.
We compare the performance of our proposed algorithm with the following baseline algorithms.

\begin{figure*}[!h]
\hspace{3mm}
\begin{minipage}[t]{0.47\linewidth} 
\includegraphics[width=75mm]{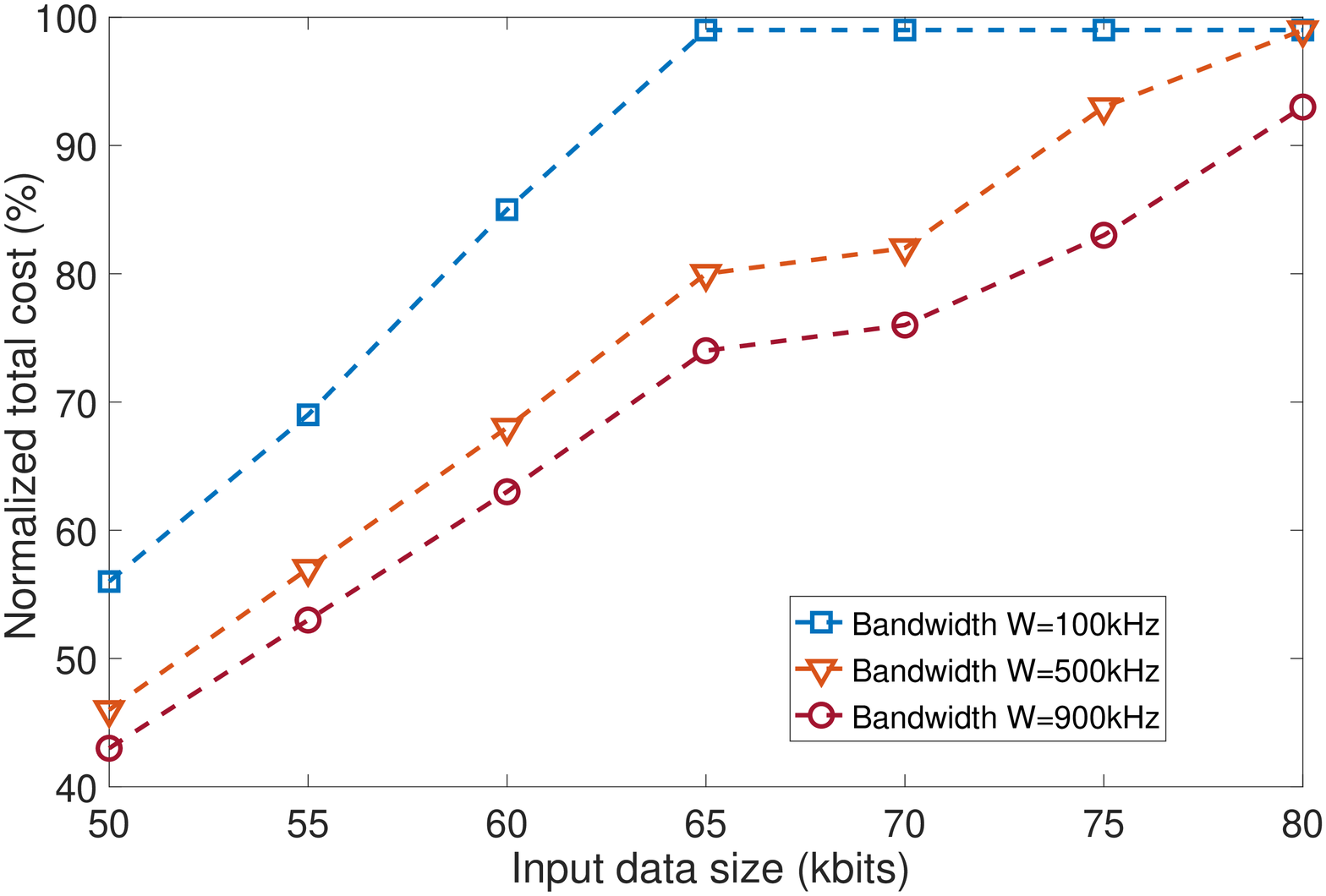}
\caption{Variations of normalized total cost with the input data size.}
\label{fig-RS_D}
\end{minipage}
\hspace{3mm}
\begin{minipage}[t]{0.47\linewidth}
\includegraphics[width=75mm]{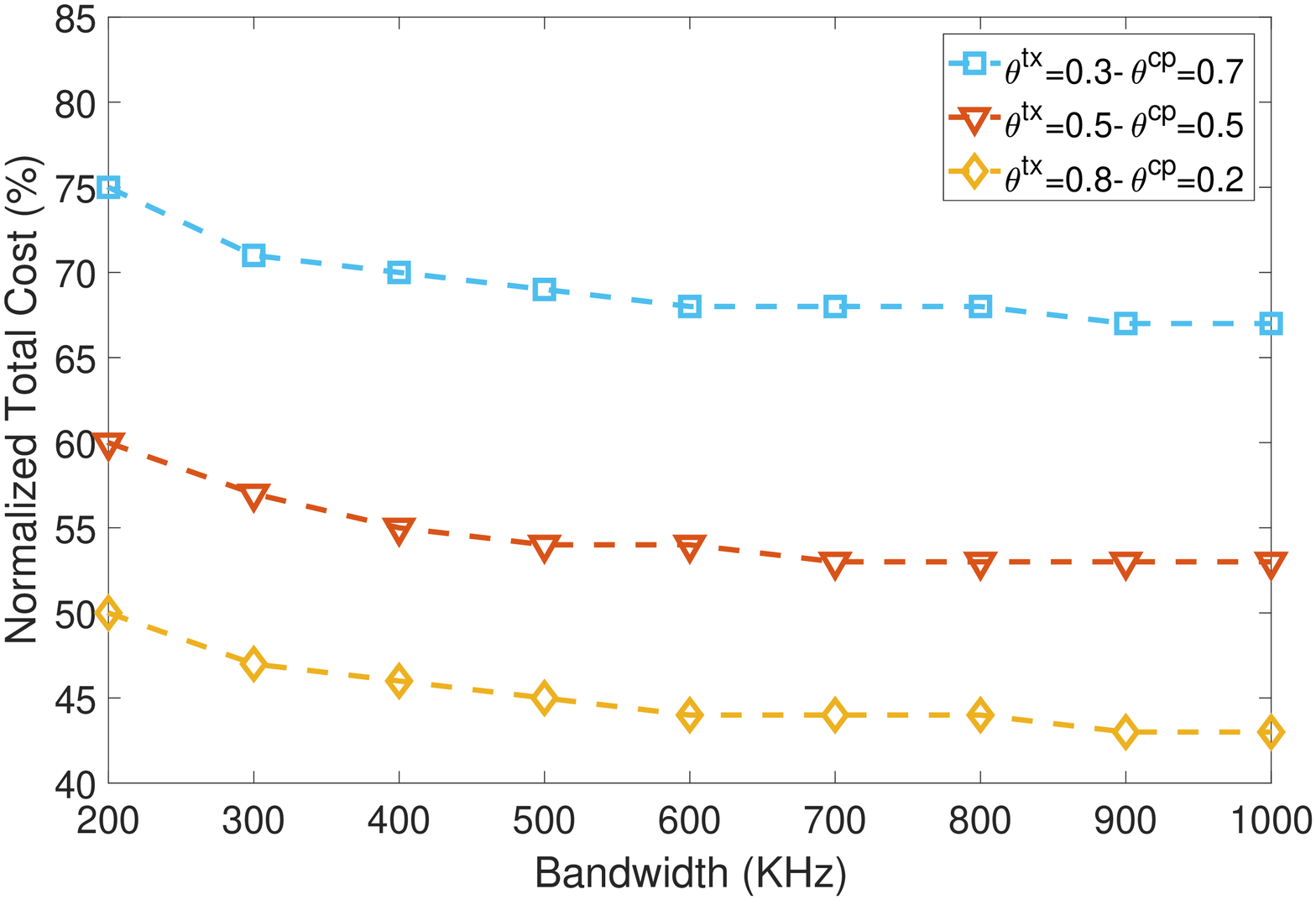}
\caption{Variations of normalized total cost with bandwidth.}
\label{fig-RS_W}
\end{minipage}
\end{figure*}

\noindent
\paragraph{Greedy Offloading and Joint Resource Allocation (\textbf{GOJRA})} In this algorithm, as many requests as
possible are offloaded up to fill up the maximum capacity of the CoSs at the BSs and then computation allocations are jointly optimized.

\noindent
\paragraph{Heuristic offloading decision algorithm (\textbf{HODA} \cite{lyu2016multiuser})}
This algorithm evaluates the cost reduction factor and
each request is offloaded if its cost reduction is positive and vice versa. The algorithm 
is run at each BS to receive all offloading requests and then jointly decides offloading requests based on 
the sign of the corresponding cost reduction factors.

First, we examine the variations of the normalized total cost (the value of the considered objective function)
 versus the input data size in Fig.~\ref{fig-RS_D}.
It can be seen that the system uses almost all system resource in the low bandwidth scenario with W=100kHz. 
Low bandwidth creates the bottleneck in the communications and this can be relaxed by allocating more bandwidth resource
(W{=}500kHz). When more  bandwidth is allocated with W${=}$1MHz, the system becomes more constrained by the computing
resources so the normalized total cost can only be reduced moderately. 
In Fig.~\ref{fig-RS_W}, we show the impacts of wireless bandwidth to the achievable system cost. 
This figure shows that the setting with  $\theta^{\mathrm{\textbf{tx}}}{=}0.8$ and  $\theta^{\mathrm{\textbf{cp}}}{=}0.2$
 achieves about 30\% reduction of the normalized total cost compared to the setting with $\theta^{\mathrm{\textbf{tx}}}{=}0.3$ and  
$\theta^{\mathrm{\textbf{cp}}}{=}0.7$. This illustrates the impacts of cost weights to the achievable performance.

\begin{figure*}[!h]
\hspace{0.1cm}
\begin{minipage}[t]{0.47\linewidth}
\includegraphics[width=75mm]{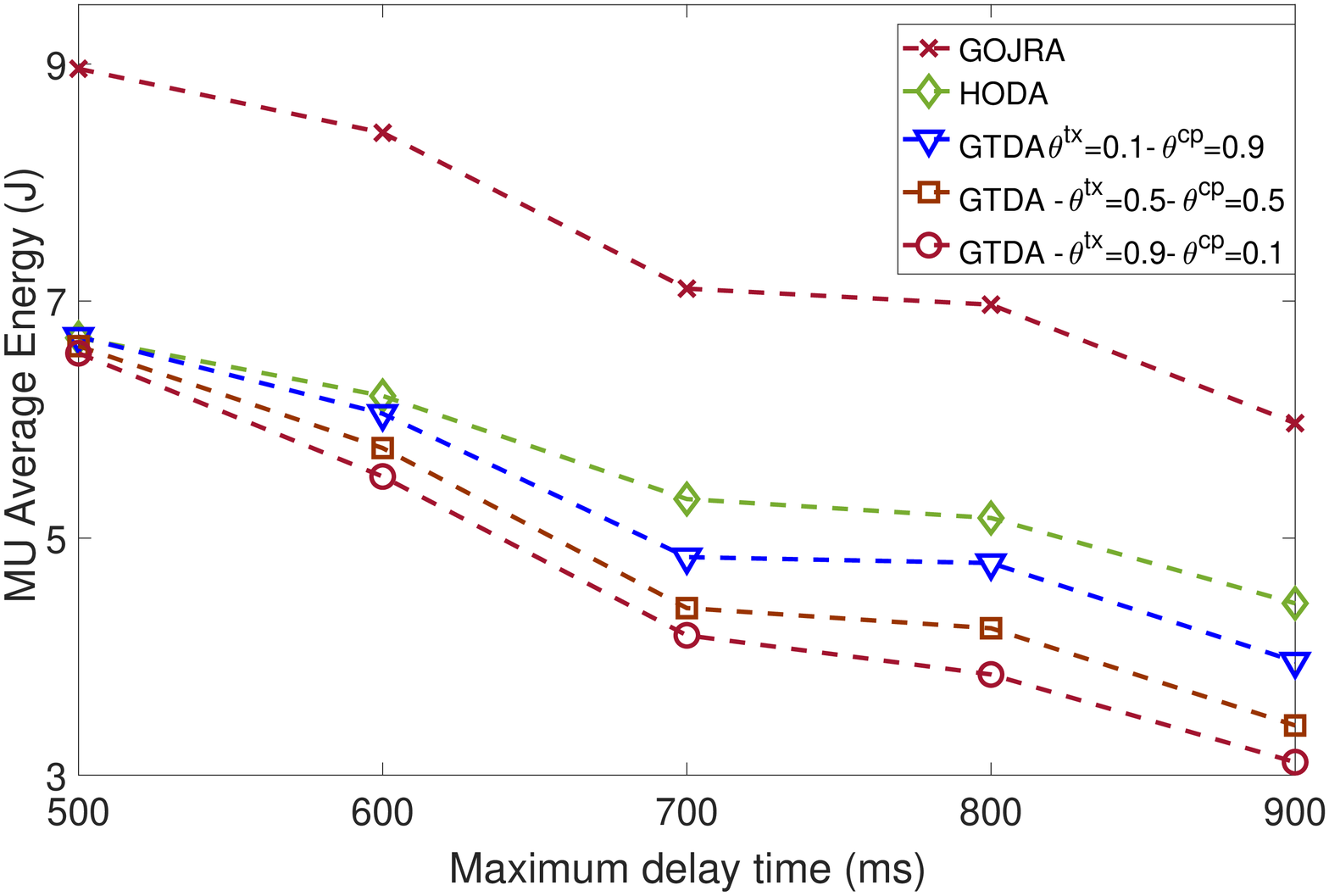}
\caption{Variations of average MU energy consumption with maximum delay.}
\label{fig-E_T}
\end{minipage}
\hspace{0.1cm}
\begin{minipage}[t]{0.47\linewidth} 
\includegraphics[width=75mm]{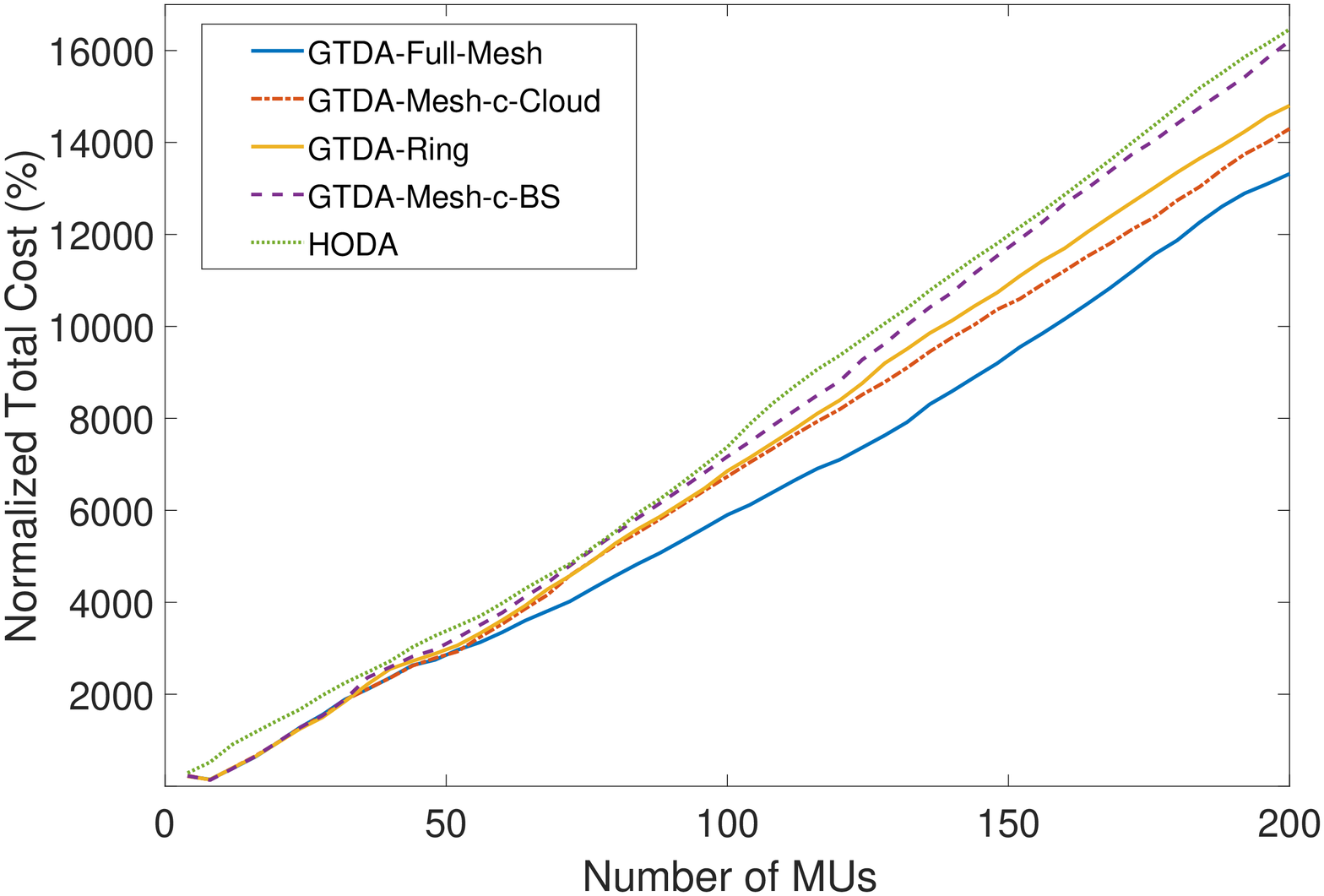}
\caption{Normalized total cost vs number of MUs.}
\label{fig-Offl_MU}
\end{minipage}
\end{figure*}

Since intelligent computation offloading can help save energy in general, we show the average energy consumption versus
the maximum allowable delay 
in Fig.~\ref{fig-E_T}. 
This figure confirms that the proposed algorithm can achieve the smallest energy among the algorithms
(i.e., GOJRA, HODA, GTDA). Moreover, the larger the allowable delay, the larger energy saving that can be achieved.

 
In Fig.~\ref{fig-Offl_MU}, we show the benefit of cooperation among the CoS where the normalized total cost 
is shown for four different network configurations: Full Mesh, Ring, Mesh-c-Cloud (with cloud servers in 
the center), and Mesh-c-BS (with the BSs' fog servers in the center) 
versus the number of MUs. The figure confirms that the Full-Mesh backhaul topology
 results in the lowest normalized total cost. This is because this topology allows
most efficient placement of functions and exploitation of computation resources.
The Mesh-c-BS topology achieves similar cost with HODA that is higher than those of other backhaul topologies.
Moreover, Mesh-c-Cloud topology leads to a slightly lower cost than that achieved by Ring topology.

\begin{figure}[!t]
\centering
\includegraphics[width=75mm]{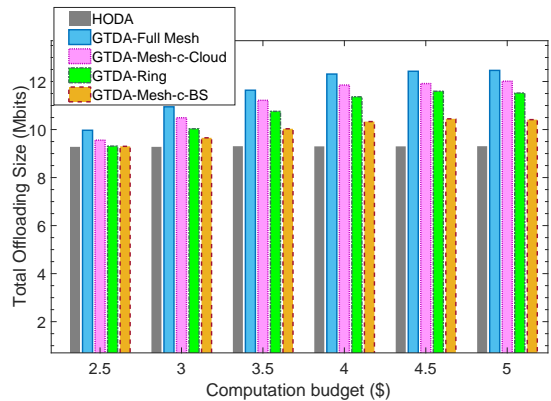}
\caption{Total offloading size vs computation budget.}
\label{fig-Offl_Bdgt}
\vspace{3mm}
\end{figure}

In Fig.~\ref{fig-Offl_Bdgt} we demonstrate the impact of the computation budget on the offloading data size associated with all offloaded 
requests considering different backhaul topologies. As can be seen, the proposed GTDA enables more effective exploitation of the computing
resources of CoSs compared to HODA. 
Moreover, the total offloading data size under the Full-Mesh topology is largest while the offloading data size for the Mesh-c-BS topology is slightly higher than that due to HODA because GTDA can leverage cooperation among BSs. 
The Mesh-c-Cloud topology leads to larger offloading data size compared to the Ring topology.
Finally, the offloading data size increases with the computation budget for all algorithms and topologies
and it becomes saturated when the computation budget is sufficiently large.

\section{Conclusion}
In this paper, we have considered the joint optimization design for the cooperative multi-server MEC system to minimize the weighted sum of MUs' energy consumption and  computing cost.
We have developed the sub-optimal
but efficient algorithm to solve the underlying problem.
Numerical results have confirmed the desirable performance
of the proposed design and the benefits of servers' cooperation. 
Specifically, the normalized total cost achieved by the
proposed algorithm is much smaller than other base line schemes.
Moreover, the full-mesh backhaul topology enables
the most efficient cooperation among CoSs and computing resource utilization; therefore,
the full-mesh backhaul topology achieves the smallest total cost compared to those achieved by other topologies.

\bibliographystyle{IEEEtran}
\bibliography{ms}

\end{document}